\documentclass[twocolumn,aps,prl,showpacs,groupedaddress]{revtex4}
\usepackage{epsfig,amssymb,amsmath}
\usepackage[hypertex,linkcolor=red]{hyperref}

\usepackage{color}

\def\labell#1{\label{#1}}
\def\section#1{{\par\em #1:--- }}
\def\togli#1{}

\begin{document}

\title{Sub-Heisenberg estimation strategies are ineffective}
\author{Vittorio Giovannetti$^1$ and Lorenzo Maccone$^2$}
\affiliation{ \vbox{ $^1$NEST, Scuola Normale Superiore and Istituto
    Nanoscienze-CNR, \\ piazza dei Cavalieri 7, I-56126 Pisa, Italy\\
    \vbox{$^2$Dip.~Fisica ``A.~Volta'', Univ.~of Pavia, via Bassi 6,
      I-27100 Pavia, Italy}} }
\begin{abstract}
  In interferometry, sub-Heisenberg strategies claim to achieve a
  phase estimation error smaller than the inverse of the mean number
  of photons employed (Heisenberg bound).  Here we show that one can
  achieve a comparable precision without performing any measurement,
  just using the large prior information that sub-Heisenberg
  strategies require. For uniform prior (i.e.~no prior information),
  we prove that these strategies cannot achieve more than a fixed gain
  of about $1.73$ over Heisenberg-limited interferometry. Analogous
  results hold for arbitrary single-mode prior distributions. These
  results extend also beyond interferometry: the effective error in
  estimating any parameter is lower bounded by a quantity proportional
  to the inverse expectation value (above a ground state) of the
  generator of translations of the parameter.
\end{abstract}
\pacs{03.65.Ta,03.67.Ac,07.60.Ly,42.50.Gy,06.20.-f,42.50.Lc}
\maketitle
It has been known for a long time that, under very general
conditions~\cite{ou,yurke,barry,luisperina,bollinger,summypegg,wiseman,HALL,HOLEVO,HAYA10-1},
the precision $\Delta\phi$ in determining the phase in interferometry
is bounded by the inverse of the mean total number of photons $N$ used
in the estimation as $\Delta\phi\geqslant 1/N$: the Heisenberg bound
to optical interferometry.  Recent work \cite{dowling,rivasluis} has
challenged the validity of such bound in regimes not covered by prior
proofs, raising the question of whether it is possible to implement
sub-Heisenberg strategies where the error has a scaling smaller
than~$1/N$.  A negative answer was given in~\cite{glm} for those
estimation strategies that are capable of reducing significantly the
initial uncertainty on the phase.  This result was obtained in the
theoretical framework of quantum estimation theory~\cite{HEL,REV1},
invoking the quantum speed limit theorem~\cite{qspeed} -- see also
Ref.~\cite{HALLETAL}. Here we show that the same negative answer
applies to essentially all estimation strategies: if sub-Heisenberg
scalings could be attained, they would be of limited use as the
overall gain in accuracy they provide is limited to a (small) constant
factor over the prior information available. To prove this, we exploit
the quantum Ziv-Zakai (ZZ) lower bound to precision introduced by
Tsang~\cite{mankei}.  Differently from the quantum Cramer-R{a}\'{o}
bound~\cite{HEL,REV1} it is not tight in general and cannot be used to
study the optimal performance of an estimation procedure, but it
provides a bound on the achievable precision that explicitly depends
on the probability distribution characterizing the prior information
of the problem. The inequalities we obtain here are an extension of
the ones of \cite{glm}, since here the prior information on the
parameter to be estimated is included.

We note that there have been prior debates on sub-Heisenberg scaling
\cite{jeff,dow,br}, and proposals to use non-linear schemes
\cite{rb,zw}: a detailed review is found in \cite{REV1}.

We start reviewing the ZZ bound~\cite{mankei} showing that, in the
regime of large information gain ({\it Low Prior Information} (LPI)
regime), it reproduces the results of Ref.~\cite{glm}. Namely, in this
regime we prove that no sub-Heisenberg scaling can be attained,
independently of the probability distribution that characterizes the
prior information on the parameter. In the opposite regime ({\it High
  Prior Information} (HPI) regime) instead, we show that, even though
sub-Heisenberg scalings are not prohibited by the ZZ bound, the
resulting accuracy is of the same order of the one obtainable by
guessing a random value distributed according to the prior
distribution (without performing any measurement). This last result is
derived under specific, but reasonable, assumptions on the prior
distribution.  In particular, we show that for uniform prior the ZZ
bound forces any sub-Heisenberg scheme to accuracies which are only a
meager $1.73$ times larger than the one obtainable via a random guess.

The final part of the paper contains material which is not directly
related with the Heisenberg bound issue, but which makes use of the
same technique developed in the previous sections.  Specifically,
generalizing previous results of~\cite{mankei}, we show how the
quantum ZZ bound can be used to derive a weighted uncertainty relation
weighted by generic prior distributions.

\section{Lower bound on the weighted precision}
Let $x$ be the parameter to be measured, say the relative optical
phase acquired by an optical pulse when traveling through the two arms
of an interferometer. We allow $x$ to take any possible real value and
assign a prior probability distribution $p(x)$ characterized by the
quantity $W$ which measures the initial uncertainty of the problem --
e.g.  for $p$ Gaussian, $W$ can be identified with the standard
deviation~\cite{NOTE}.  As is customary in quantum
metrology~\cite{REV1}, we assume $x$ to be encoded into a state
$\rho_x$ of a quantum probe system (say the optical pulse emerging
from the Mach-Zehnder interferometer) which we measure obtaining the
random outcome $y$. We can quantify the accuracy of the estimation
through the weighted Root Mean Square Error (RMSE)~\cite{mankei}:
\begin{eqnarray}
  \Delta Y:= \Big[ \int dx\:\int dy\:p(y|x)\;p(x)\;[X(y)-x]^2\Big]^{1/2}
\labell{del},
\end{eqnarray}
where $X(y)$ is the estimation of $x$ we construct from the outcome
$y$, and where $p(y|x)$ is the conditional probability distribution of
getting a certain $y$ given $x$. A lower bound for $\Delta Y$ follows
from the quantum ZZ bound~\cite{mankei},
\begin{eqnarray}
\Delta Y&\geqslant&\Big\{ \tfrac
12\int_0^{\infty}d{z}\:{z}\int_{-\infty}^{+\infty}
dx\min[p(x),p(x+z)]  \nonumber\\&& \times
\left[1-\sqrt{1-F(\rho_x,\rho_{x+z})}\right] \Big\}^{\frac{1}2}
\labell{mankei}\;,
\end{eqnarray}
where $F$ is the fidelity between the states $\rho_x$ and $\rho_{x+z}$
that correspond to a true value of the parameter of $x$ and $x+z$
respectively (an alternative, slightly stronger inequality is
presented in the Appendix). The parameter $x$ is mapped
onto the probe state $\rho_x$ by a unitary
\begin{eqnarray}
x \rightarrow \rho_x =e^{-ixH}\rho_0e^{ixH}\;,
\end{eqnarray}
with $H$ the generator of translations of $x$ (an effective
Hamiltonian).  Then, a bound for the fidelity $F$ is~\cite{qspeed,glm} 
\begin{eqnarray}\label{speed}
  F(\rho_x,\rho_{x+{z}})\geqslant\alpha^{-1}(2{\cal H}{z}/\pi)\;,
\end{eqnarray}
where ${\cal H}:=\mbox{Tr}[H\rho_0]-{\cal H}_0$ is the expectation
value of the generator $H$ above the ground level ${\cal H}_0$ of $H$,
and the function $\alpha^{-1}$ is the inverse of the function
$\alpha(\epsilon)$ defined in Ref.~\cite{qspeed} which vanishes for
$\epsilon\geqslant 1$ and is approximated by
$4$arccos$^2(\sqrt{\epsilon})/\pi^2$ for $\epsilon \in[0,1]$. In
interferometry, $H$ is the number operator of the optical
mode~\cite{REV1} and ${\cal H}$ is with the mean photon number $N$
employed in the experiment.  Replacing (\ref{speed}) into
\eqref{mankei} we find~\cite{NOTEnew}
\begin{eqnarray} 
  \Delta Y &\geqslant &  \Delta Y_{\text{\tiny{LB}}} \;, \mbox{ where}
  \label{boundnew}\\
  \Delta Y_{\text{\tiny{LB}}}  & :=& \left\{   \frac
    {x_0^2}2\int_0^{1}dt\:t\:E(x_0t )
    \left[1-\sqrt{1-\alpha^{-1}(t)}\right] \right\}^{\frac{1}2} \labell{l2}
   \end{eqnarray} 
 with $x_0:=\pi/(2{\cal H})$ and 
  \begin{eqnarray} 
  E({z})&:=&\int_{-\infty}^{+\infty}dx\min[p(x),p(x+z)]   
\labell{defe}
\;
\end{eqnarray}
being a function which satisfies the constraint $0\leqslant E\leqslant
1$, with $E(0)=1$ and $E(z)\simeq 0$ for $ z\gg W$~\cite{NOTE}. 

\section{Asymptotic regimes}
The inequality~\eqref{boundnew} is an extension of the bound given in
Ref.~\cite{glm}, as it also contains the prior information in the
function $E$. There is no guarantee that the accuracy $\Delta Y$ can
reach $\Delta Y_{\text{\tiny{LB}}}$: indeed most likely the achievable
optimal threshold for $\Delta Y$ is larger, see also~\cite{mankei}.
Still the inequality~(\ref{boundnew}) is useful to analyze whether
sub-Heisenberg scalings are possible or not.  In particular, when the
integral in \eqref{l2} can be approximated by a constant, the factor
$x_0^2$ implies a $1/{\cal H}$ scaling for $\Delta Y$ as requested by
the Heisenberg bound.  This happens for instance in the LPI regime
considered in~\cite{glm} where the Heisenberg scaling $1/{\cal H}$
guarantees a large improvement over the prior uncertainty measured by
$W$, i.e.~when the ratio $t_0:=W/x_0=2 {\cal H} W/\pi$ diverges.  In
this case in fact, independently from the specific form of the prior
distribution $p(x)$, one has $\left.
  E(x_0t)=E(Wt/t_0)\right|_{t_0\rightarrow \infty} \rightarrow 1$ for
all values of the integral in \eqref{l2} and hence
\begin{eqnarray}
\Delta Y_{\text{\tiny{LB}}}^{\text{\tiny{(LPI)}}} &:=&
\lim_{t_0\rightarrow \infty} \Delta Y_{\text{\tiny{LB}}}
= x_0 \sqrt{\frac{A}2 }\labell{asympt}
=\sqrt{\frac{ A}{2}}\;\frac{\pi}{2\cal H}\;,\\
\mbox{with }  A &:=&\int_0^{1}dt\:t[1-\sqrt{1-\alpha^{-1}(t)}] \simeq 0.042\;.
\labell{defaaa}
\end{eqnarray}
A similar result was derived in~\cite{mankei} under the assumption of
a uniform prior distribution $p(x)$, and by bounding~$\alpha^{-1}$ of
Eq.~(\ref{speed}) with a linear function.  Since the asymptotic value
$\Delta Y_{\text{\tiny{LB}}}^{\text{\tiny{(LPI)}}}$ does not depend on
the specific choice of the prior and has a Heisenberg scaling, we will
use it as a benchmark to evaluate the performances of sub-Heisenberg
strategies~\cite{NOTE11}.

Consider now the HPI regime in which the prior uncertainty is already
much better than the one we could possibly get from a Heisenberg-like
scaling. Indicatively this can be identified by the condition $t_0 \ll
1$.  Since for all $t\neq 0$, $E(x_0t)=E(Wt/t_0)$ vanishes
when $t_0\rightarrow 0$~\cite{NOTE} one expects that $\Delta
Y_{\text{\tiny{LB}}}$ of Eq.~(\ref{boundnew}) reduces drastically.
As a consequence, the bound~(\ref{boundnew}) no longer prevents $\Delta
Y$ from dropping below the $1/{\cal H}$ scaling defined by the
benchmark value of Eq.~(\ref{asympt}). However, this is hardly
surprising: in the HPI regime the real question is whether or not one
can get a significant improvement with respect to the {\it initial}
uncertainty. Drawing general conclusions on this issue is difficult
due to the complex dependence of~(\ref{l2}) from the prior probability
 $p(x)$.  Interestingly, in the case of single-mode
distributions $p(x)$ (i.e.~with a single maximum) we can show that the
improvement on $\Delta Y$ vanishes for $t_0\rightarrow 0$.  To prove
this we note that for single-mode distributions Eq.~\eqref{defe} yields
\begin{eqnarray}
E(x_0t)
&=&1-\int_{y_m-x_0{t}/2}^{y_m+x_0t/2}dx\:p(x)
\labell{l4}\;,
\end{eqnarray}
with $y_m:=y_0+ x_0t/2$, where $y_0$ is the point where
$p(y_0)=p(y_0+x_0 t)$ (such point is unique for single-mode
distributions). Remembering that the probability distribution $p(x)$
has a width $W$, from \eqref{l4} it is clear that for $t_0\to 0$ the
function $E(x_0 t)=E(Wt/t_0)$ is non-null only when $t\simeq 0$.
Hence, to gauge the lower bound \eqref{boundnew} in the HPI regime we
can remove the term in square brackets in \eqref{l2}, since
$\alpha^{-1}({t})\simeq 1$ for ${t}\simeq 0$. Accordingly,
\eqref{l2} becomes
\begin{eqnarray}
\Delta^2 Y_{\text{\tiny{LB}}}=
\tfrac{x_0^2}2\int_0^1dt\:{t}\:E(x_0{t})=\Gamma+\tfrac{x_0^3 }{8}
\int_0^1dt\:t^2[p(y_m+\tfrac{x_0{t}}{2})&&\nonumber\\+p(y_m-\tfrac{x_0{t}}{2})]
=\Gamma+ \; \int_{y_m-\tfrac{x_0}{2}}
^{y_m+\tfrac{x_0}{2}}dx\;(x-y_m)^2p(x)  \nonumber 
\;,\qquad&&
\end{eqnarray}
where the first equality follows from an integration by parts, whence
$\Gamma:=x_0^2E(x_0)/4$, and the second equality is a change of
integration variables. If the prior $p(x)$ has a variance
$\Delta^2X$ \cite{notao}, for $t_0\to 0$ the above expression is
\begin{eqnarray}
  \Delta Y_{\text{\tiny{LB}}}^{\text{\tiny{(HPI)}}}:=
  \lim_{t_0\rightarrow 0}  \Delta Y_{\text{\tiny{LB}}}  
  =
  \sqrt{\Delta^2X+(y_m-\mu)^2}
  \geqslant\Delta X ,
\labell{l6}
\end{eqnarray}
where $\mu$ is the distribution mean, and where we used the fact that
$\lim_{t_0\to\infty}\Gamma=0$~\cite{NOTE}.  The asymptotic inequality
\eqref{l6} shows that the estimation error is lower bounded by the
standard deviation $\Delta X$ of the prior distribution: the same
precision of a strategy that just takes a random guess (according to
the prior distribution) for the parameter to estimate. Therefore,
sub-Heisenberg schemes are useless in this regime: one can attain
the same precision just by ``guessing'' the parameter.

\begin{figure}[t]
\begin{center}
\epsfxsize=.4\hsize\leavevmode
\epsffile{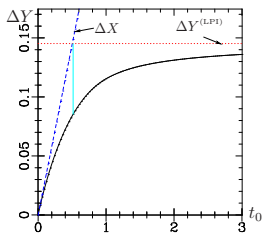}
\end{center}
\vspace{-.5cm}
\caption{Comparison between the bounds on the precision of the
  estimation for uniform prior. The continuous line is the lower bound
  $\Delta Y_{\text{\tiny{LB}}}$ defined in \eqref{l3}. The dashed line
  is the accuracy $\Delta X=W/\sqrt{12}=x_0t_0/\sqrt{12}$ one gets by
  guessing the value of $x$ from the prior distribution.  The dotted
  line is the asymptotic limit $\Delta
  Y_{\text{\tiny{LB}}}^{\text{\tiny{(LPI)}}}$ of~\eqref{asympt} which
  provides the benchmark for the Heisenberg scaling.  For $t_0> 1$ the
  bound~(\ref{boundnew}) implies that the relative gain $\Delta
  Y_{\text{\tiny{LB}}}^{\text{\tiny{(LPI)}}}/\Delta Y$ one can
  achieve with sub-Heisenberg strategies is limited by a
  factor smaller than 2. For small $t_0$ the gap increases, but in
  this case, one has to compare the performance of sub-Heisenberg
  strategies with the uncertainty $\Delta X$ of the prior which also
  reduces: the vertical segment indicates the maximum gain $\simeq
  1.73$. \togli{ The circles represent the precision achievable through the
  sub-Heisenberg scheme detailed in \cite{rivasluis}, obtained through
  a Monte-Carlo simulation and plotted versus the true value. }Here
  $x_0=1$.  }
\label{f:1}
\vspace{-.5cm}
\end{figure}
\section{Intermediate regimes} 
Having discussed the general behavior of the bound~(\ref{boundnew})
for large and small $t_0$, here we analyze some examples of prior
distributions for which the function $\Delta Y_{\text{\tiny{LB}}}$ can
be explicitly evaluated.  This allows us to study regimes which are
intermediate between  HPI and LPI, showing that sub-Heisenberg
strategies can deliver advantages which, at most, are limited to a
(small) constant factor with respect to the prior accuracy.

Probably the most interesting case is when one has no prior
information about the true value of the parameter, which corresponds
to a uniform prior $p(x)$ of width $W$ with equal weight for all
possible values~\cite{mankei}. In this case, we have $E(z)=1-z/W$ for
$z\leqslant W$, and $E(z)=0$ otherwise. Thus, Eq.~\eqref{l2} becomes
\begin{eqnarray}
  \Delta Y_{\text{\tiny{LB}}}&=&x_0 \;\sqrt{ [A(t_0)-B(t_0)/t_0]/2}\;,\labell{l3}\; \mbox{ with }\\
  A(t_0)&:=&\int_0^{\min(t_0,1)}dt\:t[1-\sqrt{1-\alpha^{-1}(t)}]\;,
\labell{defa}\\
B(t_0)&:=&\int_0^{\min(t_0,1)}dt\:t^2[1-\sqrt{1-\alpha^{-1}(t)}]
\labell{defb}\;,
\end{eqnarray}
which is plotted in Fig.~\ref{f:1}. In agreement with the analysis of
the previous section, for large $t_0\to\infty$ Eq.~\eqref{l3} yields
the universal limit~(\ref{asympt}) --- indeed $A(\infty)=A$ and
$B(\infty)/t_0\to 0$. Moreover, the trivial estimation strategy of
choosing a random value according to the prior distribution gives an
error $\Delta X=W/\sqrt{12}=x_0 t_0/\sqrt{12}$. In the HPI regime
($t_0\ll 1$) the term in square brackets in \eqref{defa} and
\eqref{defb} can again be removed since $\alpha^{-1}({t})\sim 1$ for
${t}\sim 0$.  Hence, the lower bound \eqref{l3} gives $\Delta
Y_{\text{\tiny{LB}}}={x_0}\sqrt{[t_0^2/2-t_0^2/3]/2}=W/\sqrt{12}=\Delta
X$, matching the trivial random estimation procedure in agreement
with~(\ref{l6}). Furthermore, the largest gap between the lower bound
(\ref{l3}) and the prior uncertainty is reached for $t_0\simeq
.5$ 
(vertical line in Fig.~\ref{f:1}): even assuming that a sub-Heisenberg
strategy could reach (\ref{l3}), this will only provide a relative
gain of about $1.73$ with respect to the initial uncertainty.

\begin{figure}[t]
\begin{center}
\epsfxsize=.8\hsize\leavevmode
\epsffile{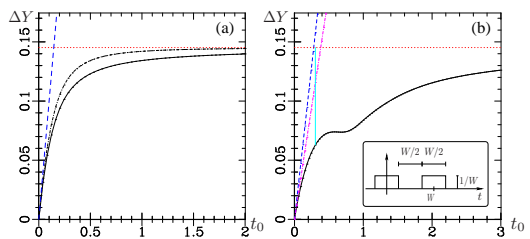}
\end{center}
\vspace{-.5cm}
\caption{Alternative priors. As in Fig.~\ref{f:1}, the continuous line
  represents the lower bound~(\ref{boundnew}), the dashed line
  represents the prior uncertainty provided by the standard deviation
  $\Delta X$ of the prior, the dotted line is the asymptotic
  Heisenberg scaling~(\ref{asympt}), and $x_0=1$.  (a) Gaussian with $W=\Delta X$:
  the bound \eqref{boundnew} here takes the form $\Delta^2Y\geqslant
  \int_0^1dt\:{t}[1-erf(\tfrac
  t{\sqrt{8}t_0})][1-\sqrt{1-\alpha^{-1}({t})}]$. The alternative
  bound \eqref{mankei3} is plotted as a dash-dotted line.  (b) The
  bimodal distribution defined in the inset, with $W=x_0t_0$ and
  $\Delta X=x_0 t_0 \sqrt{13/48}$ (dashed line) [here the asymptotic
$t_0\to 0$  behavior is $\Delta Y=x_0 t_0\sqrt{7/{48}}$ (dash-dotted line)].  }
\label{f:2}
\vspace{-.5cm}
\end{figure}

Similar results have been obtained for other prior distributions. In
particular, in Fig.~\ref{f:2} we report the results obtained for
Gaussian 
and for a bimodal step-like prior distribution. In all cases we record
maximum gains of order 2 over the prior uncertainties.

\section{Conclusions} 
Using the quantum ZZ bound recently introduced in~\cite{mankei} we
have shown that, independently of the prior information available at
the beginning of the protocol, the resulting accuracy cannot beat the
Heisenberg scaling $1/{\cal H}$ in the LPI regime (which is arguably
the most relevant for practical implementations).  Moreover, we have
shown that in the other cases where the prior is sufficiently large to
allow for a sub-Heisenberg scaling, the resulting accuracy cannot
attain a significant enhancement over the one available before
starting the estimation procedure.

A similar analysis can be done by replacing ${\cal H}$ with the
standard deviation $\Delta{\cal H}$ of the generator $H$. This yields
a generalized uncertainty relation \cite{UNCERT} weighted by the
prior: It is sufficient to replace the inequality~(\ref{speed}) with
the Bhattacharyya-like inequality~\cite{bhatta} of Ref.~\cite{qspeed},
i.e.  $F(\rho_x,\rho_{x+z})\geqslant\cos^2(\Delta {\cal H}z)$ for
$\Delta{\cal H}z\leqslant\pi/2$. Replacing this into \eqref{mankei},
we obtain
\begin{eqnarray}
  \Delta Y
  \geqslant \Big\{ \frac
  {\delta_0^2}2\int_0^{1}dt\:t\:E(\delta_0t)
  [1-\sin({t}\pi/2)] \Big\}^{\frac{1}2}
\labell{bha}\;,
\end{eqnarray}
where $\delta_0:=\pi/(2\Delta{\cal H})$.  From this it follows again
that in the asymptotic regime $t_0\to\infty$ one has $\Delta
Y\geqslant\delta_0 \sqrt{A'/2}$ where $A'=1/2-4/\pi^2\simeq 0.095$ is
a constant that does not depend on the prior. In contrast, for
$t_0\to 0$, and assuming a single-mode prior, the bound \eqref{bha}
becomes $\Delta Y\geqslant\Delta X$, showing that the initial
uncertainty again determines the scaling of the accuracy.

\section{Appendix} 
An inequality slightly stronger than~(\ref{mankei}) can be obtained by
using the quantum ZZ-bound~\cite{mankei}
\begin{eqnarray}
\Delta Y&\geqslant&\Big\{ \tfrac
12\int_0^{\infty}d{z}\:{z}\int_{-\infty}^{+\infty}
dx \; [p(x)+p(x+z)]  \nonumber\\&& \times
Pr_e(x,x+z)  \Big\}^{\frac{1}2}
\labell{mankei2}\;,
\end{eqnarray}
where $Pr_e(x,x+z)$ is the minimum error probability of a
hypothesis-testing problem which aims to discriminates between the
states $R_0:=\rho_x$ and $R_1:=\rho_{x+z}$ assuming that they are
produced with probabilities $P_0=p(x)/[p(x)+ p(x+z)]$ and $P_1=1-P_0$,
respectively.  Introducing two purifications $\Psi_0$ and $\Psi_1$ of
$R_0$ and $R_1$ respectively, which verify the identity
$F(\rho_0,\rho_1)=F(\Psi_0,\Psi_1)$, the quantity $Pr_e(x,x+z)$ can be
bounded as follows
\begin{eqnarray}
Pr_e(x,x+z) &\geqslant& \frac{1}2 \left[ 1- \| P_0 R_0 - P_1 R_1\|_1\right] \nonumber \\
&\geqslant&  \frac{1}2 \left[ 1-  \| P_0 \Psi_0 - P_1 \Psi_1\|_1\right] \nonumber\\
&=&  \frac{1}2 \left[ 1- \sqrt{ 1- 4 P_0 P_1 F(\Psi_0,\Psi_1)}  \right] \nonumber \\ 
&=&    \frac{1}2 \left[ 1- \sqrt{ 1- 4 P_0 P_1 F(R_0,R_1)}  \right] \;, \nonumber 
\end{eqnarray} 
where the first inequality derives from~\cite{HEL} (see also
Ref.~\cite{mankei}) and the second from the fact that the trace
distance is decreasing under the action of partial trace. Replacing
this into (\ref{mankei2}) and using~(\ref{speed}) we then find
\begin{eqnarray}
\Delta Y&\geqslant&\Big\{ \frac{x_0^2}4\int_0^{1}d{t}\:{t}\int_{-\infty}^{+\infty}
dx \; [p(x)+p(x+x_0t)] \labell{mankei3}\\ \nonumber&& \times  \left[ 1- \sqrt{ 1- 4 \tfrac{p(x)p(x+x_0t)}{[p(x)+p(x+x_0t)]^2} \alpha^{-1}(t)}  \right] 
 \Big\}^{\frac{1}2}\;.
\end{eqnarray}
By direct evaluation it results that at least for values of $x_0$ and
for the prior distributions analyzed in the text, (\ref{mankei3})
provides a stronger bound than (\ref{boundnew}), see Fig.~\ref{f:2}.
However, in the LPI regime \eqref{mankei3} and \eqref{boundnew} give
the same asymptotic constraint. Indeed, since $p(x+x_0t)=p(x+Wt/t_0)$,
it follows that for $t_0\rightarrow \infty$, $p(x+x_0t) \rightarrow
p(x)$ and thus the rhs of (\ref{mankei3}) converges to the quantity
$\Delta Y_{\text{\tiny{LB}}}^{\text{\tiny{(LPI)}}}$ of
Eq.~(\ref{asympt}).

We thank M.~Tsang for useful discussions. VG acknowledges support from
MIUR through the FIRB-IDEAS Project No.~RBID08B3FM.


\begin{references}
\bibitem{NOTE} Operationally $W$ is defined by the requirement that
  $p(x)$ vanishes for $|x|\gg W$. In particular, if the distribution
  $p(x)$ possesses a variance this implies that the function $E(z)$
  defined in Eq.~(\ref{defe}) nullifies when~$z\gg W$.
\bibitem{NOTEnew}
In writing Eq.~(\ref{l2}) 
 we used the fact that $\alpha^{-1}({t})=0$ for ${t}\geqslant 1$
to choose the upper integration limit. 
\bibitem{NOTE11} Since $E$ is always smaller than 1, the lower bound
  in \eqref{boundnew} can never be larger than the asymptotic value
  ${x_0}\sqrt{A/2}$.
\end{references}
\end{document}